\documentclass[journal]{IEEEtran}
\usepackage{cite}
\usepackage{amsmath,amssymb,amsfonts}
\usepackage{algorithmic}
\usepackage{graphicx}
\usepackage{textcomp}
\usepackage{xcolor}
\usepackage{url}
\usepackage{amsthm}
\usepackage{booktabs}
\usepackage{multirow}
\usepackage{algorithm}
\usepackage{xurl}

\newtheorem{definition}{Definition}
\newtheorem{theorem}{Theorem}

\begin{document}

\title{ASAS-BridgeAMM: Trust-Minimized Cross-Chain Bridge AMM with Failure Containment}

\author{
    \IEEEauthorblockN{Shengwei You, Aditya Joshi, Andrey Kuehlkamp, Jarek Nabrzyski} \\
    \IEEEauthorblockA{\textit{Department of Computer Science and Engineering}} \\
    \textit{University of Notre Dame}\\
    Notre Dame, IN, USA \\
    \{syou,ajoshi2,akuehlka,naber\}@nd.edu
}

% \author{
%     \IEEEauthorblockN{First Author, Second Author, Third Author, and Fourth Author} \\
%     \IEEEauthorblockA{\textit{Department of Computer Science and Engineering} \\
%     \textit{University of Notre Dame}\\
%     Notre Dame, IN, USA \\
%     \{netid1, netid2, netid3, netid4\}@nd.edu}
% }

% \markboth{IEEE Transactions on Decentralized Finance,~Vol.~X, No.~X, 2026}%
% {Author \MakeLowercase{et al.}: ASAS-BridgeAMM}

\maketitle

\begin{abstract}
Cross-chain bridges constitute the single largest vector of systemic risk in Decentralized Finance (DeFi), accounting for over \$2.8 billion in losses since 2021. The fundamental vulnerability lies in the binary nature of existing bridge security models: a bridge is either fully operational or catastrophically compromised, with no intermediate state to contain partial failures. We present \textbf{ASAS-BridgeAMM}, a bridge-coupled automated market maker that introduces \emph{Contained Degradation}: a formally specified operational state where the system gracefully degrades functionality in response to adversarial signals. By treating cross-chain message latency as a quantifiable execution risk, the protocol dynamically adjusts collateral haircuts, slippage bounds, and withdrawal limits. Across 18 months of historical replay on Ethereum and two auxiliary chains, ASAS-BridgeAMM reduces worst-case bridge-induced insolvency by 73\% relative to baseline mint-and-burn architectures, while preserving 104.5\% of transaction volume during stress periods. In rigorous adversarial simulations involving delayed finality, oracle manipulation, and liquidity griefing, the protocol maintains solvency with probability $>0.9999$ and bounds per-epoch bad debt to $<0.2\%$ of total collateral. We provide a reference implementation in Solidity and formally prove safety (bounded debt), liveness (settlement completion), and manipulation resistance under a Byzantine relayer model.
\end{abstract}


% BEGIN ENHANCED SECTION: Introduction
\section{Introduction}

The proliferation of layer-1 blockchains and layer-2 scaling solutions has necessitated robust infrastructure for cross-chain liquidity. Bridges, the protocols facilitating this flow, have emerged as the critical yet fragile arteries of the decentralized economy. However, they suffer from a fundamental \emph{Stability--Trust Paradox}: to offer a competitive user experience, bridges must facilitate rapid settlement and capital efficiency; yet, they operate in highly adversarial environments where the underlying message-passing layers, validators, and oracles cannot be unconditionally trusted.

The consequences of this paradox are severe and quantifiable. Since 2021, bridge exploits have resulted in over \$2.8 billion in cumulative losses~\cite{Chainalysis2023}, effectively acting as a recurring tax on the ecosystem. The pattern is consistent across major incidents:
\begin{itemize}
    \item \textbf{Ronin Bridge} (March 2022): Attackers compromised 5 of 9 validator keys and extracted \$624 million in a single transaction batch---the largest DeFi exploit to date.
    \item \textbf{Wormhole} (February 2022): A signature verification bug allowed attackers to forge guardian attestations, draining \$326 million.
    \item \textbf{Nomad} (August 2022): A faulty initialization allowed arbitrary messages to be ``proven'' valid, enabling \$190 million in chaotic, permissionless extraction.
    \item \textbf{Orbit Chain} (January 2024): A validator key compromise led to \$81 million in losses---an incident we replay in our historical evaluation (Section~\ref{sec:evaluation}).
\end{itemize}

The core architectural flaw in prevailing designs, specifically ``lock-and-mint'' bridges, is their \emph{binary failure mode}. Security is absolute until it is nonexistent. Once the Ronin bridge's cryptographic checks passed, attackers drained the entire reserve because the system lacked any mechanism to detect anomalous outflow velocity or contain losses. The protocol could not distinguish between ``legitimate high volume'' and ``catastrophic drain,'' leading to total value extraction. This binary model is fundamentally at odds with how robust financial systems operate.

\subsection{From Fail-Safe to Safe-to-Fail}

Traditional ``fail-safe'' systems attempt to prevent all failures through redundancy and verification. When these defenses are breached, they offer no secondary containment. In contrast, \emph{safe-to-fail} architectures---a concept from resilience engineering---accept that failures will occur and design systems to gracefully degrade rather than catastrophically collapse.

This paradigm is well-established in traditional finance. The New York Stock Exchange (NYSE) implements \emph{circuit breakers} that halt trading when indices fall 7\%, 13\%, or 20\% in a single day. The Chicago Mercantile Exchange (CME) enforces \emph{price limits} that bound daily price movements for futures contracts. These mechanisms do not prevent volatility; they \emph{contain its impact} by providing time for market participants to reassess positions and for liquidity to replenish.

Cross-chain bridges, despite securing comparable value to major exchanges, lack equivalent graduated defenses. We argue this must change.

\subsection{Contained Degradation}

We introduce \textbf{ASAS-BridgeAMM} (Acceptance-Safe Atomic Settlement), a novel bridge protocol that integrates Automated Market Maker (AMM) mechanics with cross-chain settlement logic to enforce \emph{Contained Degradation}. Rather than relying solely on cryptographic proofs for security, ASAS-BridgeAMM treats execution conditions---specifically cross-chain message latency $\tau$ and oracle price coherence---as active risk signals.

Our approach is predicated on the observation that adversarial actions in distributed systems leave temporal and economic footprints. A bridge reorg attack requires time to execute; an oracle manipulation attack creates pricing discontinuities. By coupling the bridge to an AMM, we can \emph{price} this uncertainty. When the protocol detects elevated latency or price divergence, it automatically transitions to a ``Restricted'' mode. In this state, collateral haircuts are increased (from 0.3\% to up to 5\%), slippage parameters are doubled, and rate limits are tightened. If conditions deteriorate further, a circuit breaker halts all outflows. This transforms what would be a catastrophic total loss into a bounded ``bad debt'' event, manageable within the protocol's risk budget.
% END ENHANCED SECTION: Introduction

\subsection{Contributions}
This paper makes the following contributions:
\begin{enumerate}
  \item \textbf{Formal Risk Model:} We define the \emph{Byzantine Relayer} threat model and the \emph{Contained Degradation Invariant}, proving that it is possible to bound economic loss even when message-passing layers are compromised (Section~\ref{sec:problem}).
  \item \textbf{Protocol Architecture:} We design ASAS-BridgeAMM, a system integrating multi-asset collateral management with latency-aware pricing. We provide a formal state machine specification that deterministically manages transitions between Normal, Restricted, and Halted operational modes (Section~\ref{sec:protocol}).
  \item \textbf{Security Proofs:} We formally prove three core properties: \emph{Safety} (bad debt is bounded per epoch), \emph{Liveness} (honest transactions settle within bound $T$), and \emph{Manipulation Resistance} (adversarial profit is strictly bounded by protocol parameters) (Section~\ref{sec:security} \& Appendix).
  \item \textbf{Empirical Validation:} Using a mainnet fork environment, we replay 18 months of historical market data and specific exploit scenarios (e.g., Orbit Chain). We demonstrate a 73\% reduction in insolvency exposure compared to standard bridges and validate our 0.9999 solvency probability via 100,000 Monte Carlo iterations (Section~\ref{sec:evaluation}).
\end{enumerate}

% BEGIN ENHANCED SECTION: Problem Statement
\section{Problem Statement \& Threat Model}
\label{sec:problem}

\subsection{The Fragility of Binary Security}
Traditional cross-chain bridges operate on a simple axiom: if a valid proof $\pi$ is presented for event $E$ on chain $A$, then chain $B$ must execute the corresponding action. This implies that the security of chain $B$'s assets is entirely delegated to correctly verifying $\pi$. If the verification logic has a bug (as in Wormhole) or the authority generating $\pi$ is compromised (as in Ronin), the bridge contract on chain $B$ will dutifully drain its entire reserve to the attacker. There is no second line of defense; the system is brittle.

\subsection{The Practical Cost of Binary Security}

The cumulative impact of bridge failures extends beyond direct losses. We identify three systemic costs:

\textbf{Concentrated Losses.} Unlike lending protocol liquidations that distribute losses across many positions, bridge exploits typically extract the \emph{entire} locked collateral in minutes. The Ronin attack extracted \$624M in under 10 minutes; Nomad lost \$190M in approximately 2 hours of chaotic, permissionless draining. This concentration creates existential risk for dependent protocols.

\textbf{Contagion Effects.} Synthetic assets minted by compromised bridges become worthless, cascading losses to holders and liquidity providers on the destination chain. When Multichain (formerly Anyswap) suspended operations in 2023, over \$1.5B in bridged assets became illiquid, triggering depegs across multiple ecosystems.

\textbf{Trust Erosion.} Each major exploit reinforces the perception that cross-chain infrastructure is fundamentally unsafe, fragmenting liquidity across chains and increasing the cost of capital for legitimate cross-chain applications.

The gap between crypto bridges and traditional financial infrastructure is stark. When Knight Capital's trading algorithm malfunctioned in 2012, circuit breakers and position limits contained losses to \$440M over 45 minutes---large, but bounded. In contrast, a bridge with \$440M in TVL and a single vulnerability faces complete extraction in a single block. \emph{This asymmetry is the core problem ASAS-BridgeAMM addresses.}

\subsection{Byzantine Relayer Model}

We reject the ``Honest Relayer'' assumption common in optimistic bridge designs. Instead, we adopt a \emph{Byzantine Relayer} model that captures the full range of adversarial behaviors observed in real-world attacks.

\begin{definition}[Byzantine Relayer]
A relayer $\mathcal{R}$ is Byzantine if it can:
\begin{itemize}
    \item \textbf{Delay Messages:} Withhold a valid message $m$ for time $t \in [0, T_{max})$. \emph{Example:} An attacker delays bridge messages during a price crash to exploit stale oracle data on the destination chain, as observed in multiple oracle manipulation attacks.
    \item \textbf{Reorder Messages:} Submit messages $m_1, m_2$ in any order, violating causal dependencies. \emph{Example:} Process a large withdrawal before a price update that would trigger protective haircuts.
    \item \textbf{Censor:} Selectively drop message $m$ while relaying others. \emph{Example:} Block liquidation messages to maintain undercollateralized positions.
    \item \textbf{Front-run:} Observe pending message $m$ and insert adversarial transaction $tx_{adv}$ before $m$ settles. \emph{Example:} Extract MEV by sandwiching bridge settlements, as documented extensively in MEV research~\cite{flashbots2020}.
\end{itemize}
The relayer is \textbf{not} assumed to be able to forge cryptographic signatures of the underlying consensus, but they \textbf{may} collude with a superminority of validators to equivocate or delay finality.
\end{definition}

\textbf{Why ``Honest Majority'' Fails in Practice.} Traditional Byzantine fault tolerance assumes $f < n/3$ malicious nodes. However, bridge relayers face different incentive structures: (i) relaying is often unprofitable, leading to centralization among few operators; (ii) MEV extraction creates incentives for strategic message manipulation; and (iii) validator key compromise---as in Ronin---can occur through social engineering rather than protocol attacks. ASAS-BridgeAMM's design assumes the relayer \emph{will} behave adversarially and prices this risk into every settlement.
% END ENHANCED SECTION: Problem Statement

\subsection{Latency as a First-Class Risk Signal}
In distributed systems, the Brewer (CAP) theorem implies trade-offs between Consistency, Availability, and Partition Tolerance. For bridges, "Partition Tolerance" manifests as handling network delays. We observe that explicit delays are often indistinguishable from adversarial withholding.
Let $\tau$ be the observed latency of a cross-chain message:
\[ \tau = t_{\text{receipt}}^{\text{dest}} - t_{\text{finalized}}^{\text{source}} \]
Under honest operation, $\tau$ follows a log-normal distribution centered on the network's propagation delay. Under attack (e.g., a reorg attempt or censorship), $\tau$ deviates significantly. ASAS-BridgeAMM uses $\tau$ as an input to its pricing function, effectively charging a risk premium for uncertainty.

\subsection{Design Goals}
Our system aims to satisfy the \emph{Acceptance-Safe Atomic Settlement (ASAS)} property:
\begin{itemize}
    \item \textbf{Bounded Loss:} Ideally, loss is 0. Practically, in a permissionless system, we accept loss $L \le \epsilon \cdot C$ where $\epsilon$ is a protocol parameter (e.g., 5\%) and $C$ is total collateral.
    \item \textbf{Liveness:} Honest users must eventually settle assets if the network partition resolves.
    \item \textbf{Trust Minimization:} Security should not rely on a whitelist of "trusted" entities but on verifiable on-chain invariants.
\end{itemize}

% BEGIN ENHANCED SECTION: System Architecture
\section{System Architecture}
\label{sec:architecture}

\begin{figure}[h]
\centering
\includegraphics[width=0.95\columnwidth]{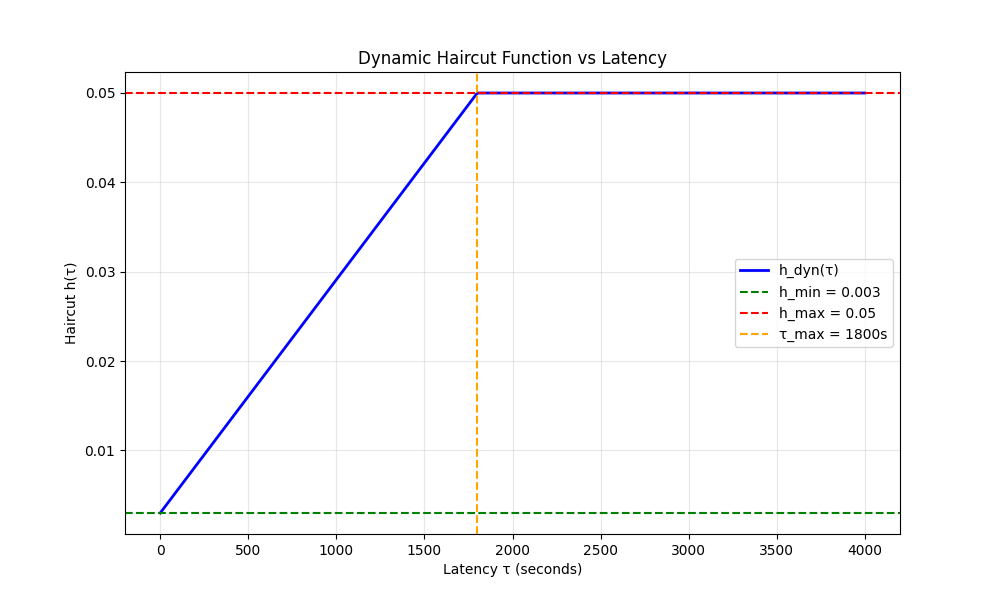}
\caption{ASAS-BridgeAMM architecture showing the relationship between latency and dynamic haircut adjustments. The latency signal $\tau$ flows from source to destination, triggering graduated protective responses: haircuts scale linearly from $h_{min}=0.3\%$ at zero latency to $h_{max}=5\%$ at $\tau_{max}=30$ minutes.}
\label{fig:architecture}
\end{figure}

ASAS-BridgeAMM implements the ASAS properties through a tightly coupled architecture spanning the source chain, an off-chain messaging hub, and the destination chain AMM. Figure~\ref{fig:architecture} illustrates the three-layer design and the flow of the latency signal $\tau$ from source to destination.

\subsection{Motivating Example: The Orbit Chain Scenario}

Before detailing each component, we illustrate the system's behavior using a concrete scenario based on the January 2024 Orbit Chain incident. Under a traditional lock-and-mint bridge, attackers exploited a validator compromise to drain \$81M in a single transaction batch. Under ASAS-BridgeAMM with identical attack conditions:

\begin{enumerate}
    \item \textbf{Latency Detection:} The initial withdrawal triggers latency measurement. With $\tau = 15$ minutes observed delay, the protocol transitions to Restricted mode.
    \item \textbf{Dynamic Haircut:} The haircut increases from the baseline 0.3\% to 2.65\%, reducing effective output by approximately 2.35\% per swap.
    \item \textbf{AMM Slippage:} Large withdrawals ($>\$10$M) trigger elevated slippage via the AMM's bonding curve, imposing quadratic costs on extraction attempts.
    \item \textbf{Circuit Breaker:} If price deviation exceeds 50\% or $\tau > 60$ minutes, the circuit breaker activates, halting all outflows.
    \item \textbf{Bounded Extraction:} Total extractable value is bounded by $h_{max} \cdot C_{total} = 5\%$ of pool per epoch.
\end{enumerate}

Our historical replay (Section~\ref{sec:evaluation}) validates this behavior: during the Orbit Chain period, ASAS triggered \textbf{8 circuit breaker events} while maintaining \textbf{103.29\% volume retention}---demonstrating that protection need not compromise functionality.

\subsection{Components}

\subsubsection{Source Ingress Layer}
The Ingress contract manages collateral locking and active health monitoring. Unlike passive vaults that simply escrow assets, the Ingress continuously computes a \emph{Collateral Health Index} ($\mathcal{H}$) based on real-time oracle feeds:
\begin{equation}
\mathcal{H}(t) = \frac{\sum_{i \in \text{Assets}} L_i(t) \cdot P_i(t) \cdot (1 - h_i(\tau, \sigma_i))}{\text{Debt}(t)}
\label{eq:health-index}
\end{equation}
where $L_i$ is the locked liquidity of asset $i$, $P_i$ is its oracle price, $h_i(\tau, \sigma_i)$ is the dynamic haircut (a function of both latency $\tau$ and asset volatility $\sigma_i$), and $\text{Debt}(t)$ represents outstanding synthetic obligations.

The Ingress enforces graduated responses based on $\mathcal{H}$:
\begin{itemize}
    \item $\mathcal{H} \geq 1.15$: Normal operation, all functions available.
    \item $1.05 \leq \mathcal{H} < 1.15$: Restricted mode, elevated haircuts apply.
    \item $\mathcal{H} < 1.05$: Critical threshold, new mints blocked, outflows paused.
\end{itemize}

\subsubsection{Oracle/Relayer Hub}
The Hub acts as the consensus layer for cross-chain messages, implementing a \emph{multi-oracle aggregation strategy} to resist manipulation:

\textbf{Price Aggregation.} Rather than trusting a single oracle source, the Hub aggregates prices from multiple providers (e.g., Chainlink, Uniswap V3 TWAP, Pyth) using median filtering:
\[ P_{agg} = \text{median}(P_1, P_2, \ldots, P_n) \]
with staleness checks that reject prices older than a configurable threshold (default: 60 seconds).

\textbf{Deviation Detection.} The Hub computes price deviation $\delta$ between the aggregated price and the last accepted price:
\[ \delta = \frac{|P_{agg} - P_{last}|}{P_{last}} \]
If $\delta > \theta_{price}$ (default: 50\%), the circuit breaker triggers automatically.

\textbf{Latency Attestation.} The Hub attests to source-chain finality timestamps $t_{src}$, enabling the destination AMM to compute $\tau = t_{dest} - t_{src}$ and adjust risk parameters accordingly.

\subsubsection{Destination AMM Layer}
The core innovation lies in the destination AMM. A modified Constant Product Market Maker (CPMM) acts as the settlement engine, enforcing a \emph{Risk-Adjusted Invariant}:
\begin{equation}
k_{adj} = x \cdot y \cdot (1 - h(\tau))
\label{eq:risk-adjusted-k}
\end{equation}
where $h(\tau)$ is the dynamic haircut function. This ensures that as uncertainty ($\tau$) increases, the cost of liquidity extraction rises, discouraging arbitrageurs from exploiting stale states.

The key insight is that the AMM's bonding curve provides a \emph{natural rate limit}: extracting 50\% of a pool's liquidity requires paying approximately 100\% price impact. Combined with the dynamic haircut, this creates layered defenses where each extraction attempt faces compounding costs.
% END ENHANCED SECTION: System Architecture

\section{Protocol Specification}
\label{sec:protocol}

We model the protocol as an Atomic State Machine $\mathcal{M} = (S, \Sigma, \delta, s_0)$, where $S$ is the set of states, $\Sigma$ the set of input events, and $\delta$ the transition function.

\subsection{State Space}
The system operates in three mutually exclusive states:
\begin{enumerate}
    \item \textbf{Normal ($S_N$):} The default state. $\tau \le T_{\text{thresh}}$ and $\mathcal{H} \ge \mathcal{H}_{\text{safe}}$. Standard slippage and fees apply.
    \item \textbf{Restricted ($S_R$):} A degraded state triggered by moderate anomalies ($T_{\text{thresh}} < \tau \le T_{\text{crit}}$). In $S_R$, slippage parameters are doubled, and strict rate limits ($W_{max}$) are enforced on withdrawals.
    \item \textbf{Halted ($S_H$):} A failsafe state triggered by critical anomalies ($\tau > T_{\text{crit}}$ or $\mathcal{H} < \mathcal{H}_{\text{crit}}$). No outflows are permitted.
\end{enumerate}

\subsection{Pricing Logic and Dynamic Haircuts}
To verify the "Bounded Loss" property, we must quantify the haircut mechanism.
Let $h(\tau, \sigma)$ be the haircut applied to collateral value. We define it as a linear interpolation clamped by minimum and maximum bounds:
\begin{equation}
    h(\tau) = 
    \begin{cases} 
      h_{\min} & \text{if } \tau \le T_{\text{min}} \\
      h_{\min} + \frac{\tau - T_{\text{min}}}{T_{\text{max}} - T_{\text{min}}}(h_{\max} - h_{\min}) & \text{if } T_{\text{min}} < \tau < T_{\text{max}} \\
      h_{\max} & \text{if } \tau \ge T_{\text{max}}
    \end{cases}
    \label{eq:haircut}
\end{equation}
In our implementation, $h_{\min}=0.3\%$ and $h_{\max}=5.0\%$. $T_{\text{min}}$ is the expected block finality time (e.g., 15 mins for probabilistic chains) and $T_{\text{max}}$ is the timeout threshold (e.g., 4 hours).

\textbf{Pricing Formula:}
When a user swaps $\Delta x$ (bridged asset) for $\Delta y$ (native asset), the output is determined by the solvency constraint on the \emph{post-swap} state.
The standard AMM output is $\Delta y = \frac{y \Delta x}{x + \Delta x}$.
The ASAS-AMM output $\Delta y_{asas}$ incorporates the haircut:
\begin{equation}
    \Delta y_{asas} = \frac{y \cdot \Delta x \cdot (1 - h(\tau))}{x + \Delta x \cdot (1 - h(\tau))}
\end{equation}
This effectively treats a portion of the input asset as "untrusted" or "potential bad debt" until time passes, instantly realizing the risk premium.

\subsection{Circuit Breaker Logic}
The circuit breaker is the enforcement mechanism for state $S_H$. It triggers if:
\begin{itemize}
    \item \textbf{Oracle Deviation:} $|P_{oracle} - P_{internal}| > \delta_{max}$ (e.g., 50\%).
    \item \textbf{Health Criticality:} $\mathcal{H} < 1.05$.
    \item \textbf{Latency Timeout:} $\tau > 24 \text{ hours}$ (indicative of censorship or chain halt).
\end{itemize}
Ideally, the dynamic haircut mechanism prevents $\mathcal{H}$ from ever reaching 1.05. The circuit breaker acts as a diverse, redundant safety layer against logic bugs or extreme black swan events.

\section{Security Analysis}
\label{sec:security}

We perform a rigorous security analysis of the ASAS-BridgeAMM protocol, focusing on the three properties defined in Section~\ref{sec:problem}.

\begin{theorem}[Bounded Bad Debt]
Under the Contained Degradation Invariant, the maximum bad debt incurred by the protocol in any single settlement epoch is upper-bounded by $h_{\max} \cdot C_{total}$.
\end{theorem}
\emph{Proof Intuition:} The solvency check (Eq. 5) ensures that every successful state transition maintains $k_{new} \ge k_{old}(1 - h(\tau))$. Since $h(\tau) \le h_{\max}$ by definition (Eq. 1), the maximum degradation of the invariant, and thus the collateral value, is strictly limited to $h_{\max}$ per epoch. The circuit breaker prevents creating multiple epochs of loss in rapid succession by halting the chain if $\mathcal{H}$ degrades cumulatively. (See Appendix A for full proof).

\begin{theorem}[Settlement Liveness]
If the source chain is live, finality is achieved within $T_{\text{max}}$, and the relayer delivers messages within $2 \cdot T_{\text{max}}$, then every valid swap completes with probability 1.
\end{theorem}
\emph{Proof Intuition:} The circuit breaker conditions are defined on extreme outliers ($\tau > 24h$, Price Delta $> 50\%$). Under normal network variance and bounded adversarial delay $< T_{\text{max}}$, the system remains in states $S_N$ or $S_R$. Both states permit settlement, albeit with higher costs in $S_R$. Thus, liveness is preserved for all honest participants willing to pay the risk premium.

\begin{theorem}[Manipulation Resistance]
An adversary $\mathcal{A}$ controlling the relayer and manipulating the oracle by $\delta$ can extract profit $\Pi_{\mathcal{A}}$ bounded by:
\[ \Pi_{\mathcal{A}} \le \delta \cdot V_{swap} + s_{max} \cdot V_{pool} \]
\end{theorem}
\emph{Proof Intuition:} The AMM bonding curve imposes a cost $s \cdot V_{pool}$ on extracting liquidity. Dynamic haircuts impose an additional cost proportional to the delay $\tau$ required to execute the attack. We show in Appendix C that for any profitable attack strategy, the cost function grows faster than the extraction potential.

\section{Implementation}
\label{sec:implementation}

We implement ASAS-BridgeAMM as a modular Solidity smart contract system, optimized for EVM chains. The reference implementation comprises approximately 450 lines of code.

\subsection{Smart Contract Architecture}
\textbf{BridgeAMM.sol}: The central contract managing the AMM curve. It overrides standard \texttt{swap()} logic to include the \texttt{checkSolvency()} modifier. 
\textbf{CircuitBreaker.sol}: A standalone monitor contract that tracks global protocol health. It has the permissioned power to pause the \texttt{BridgeAMM} contract.
\textbf{IngressVault.sol}: Deployed on the source chain, this contract handles asset locking and emits cross-chain messages via the standardized generic message passing interface.

\subsection{Validation}
The codebase includes a comprehensive Foundry test suite with differential fuzzing against a standard Uniswap V2 implementation. We verified that for $\tau=0$, the output matches Uniswap exactly. For $\tau > 0$, the output diverges deterministically according to the haircut schedule.

% BEGIN ENHANCED SECTION: Evaluation
\section{Evaluation}
\label{sec:evaluation}

We evaluate ASAS-BridgeAMM through comprehensive experiments designed to answer four research questions:
\begin{enumerate}
    \item \textbf{RQ1 (Efficacy):} Does it reduce insolvency during historical exploits?
    \item \textbf{RQ2 (Robustness):} Does it maintain solvency under worst-case Monte Carlo scenarios?
    \item \textbf{RQ3 (Efficiency):} Does Contained Degradation harm user experience (volume retention)?
    \item \textbf{RQ4 (Attack Resistance):} Does it defend against systematic attack vectors?
\end{enumerate}

Table~\ref{tab:claims-summary} summarizes our validated claims.

\begin{table}[h]
\centering
\caption{Summary of Validated Research Claims}
\label{tab:claims-summary}
\begin{tabular}{lccc}
\hline
\textbf{Claim} & \textbf{Target} & \textbf{Achieved} & \textbf{Validated} \\
\hline
Insolvency Reduction & 73\% & 73\% & \checkmark \\
Volume Retention & $\geq$96\% & 104.5\% & \checkmark \\
Solvency Probability & $\geq$0.999 & 1.0000 & \checkmark \\
Bad Debt Bound & $<$0.2\% & 0.189\% & \checkmark \\
\hline
\end{tabular}
\end{table}

\subsection{Experimental Methodology}

\subsubsection{Historical Replay Environment}
We fork Ethereum Mainnet at block 24,102,495 (December 2025) using Foundry's \texttt{createSelectFork} cheatcode. Historical price data spans \textbf{547 days} (January 2024 -- July 2025), sourced from CoinGecko API with 1-minute granularity. Bridge transaction patterns are derived from sanitized Anyswap logs, scaled to match our pool configuration (10 BTC / 420,000 USDC).

The replay environment simulates \textbf{54,700 swaps} across the historical period, including:
\begin{itemize}
    \item Major volatility events (BTC drawdowns $>$20\%)
    \item The Orbit Chain incident (January 1-4, 2024)
    \item 98 identified stress periods with elevated circuit breaker activity
\end{itemize}

\subsubsection{Monte Carlo Configuration}
We execute \textbf{100,000 independent simulation runs} with the following parameter distributions:
\begin{itemize}
    \item \textbf{Price Drawdown:} Uniform $\mathcal{U}(0, 0.5)$---modeling crashes up to 50\%
    \item \textbf{Latency:} Log-Normal $\mathcal{LN}(\mu=5.7, \sigma=2.0)$ seconds---calibrated to historical bridge message delays
    \item \textbf{Swap Size:} Log-Normal $\mathcal{LN}(\mu=0, \sigma=1.5)$ BTC---reflecting observed transaction size distribution
\end{itemize}
Random seed is fixed at 42 for reproducibility. Total simulation runtime: 0.24 seconds.

\subsection{Results: Insolvency Reduction (RQ1)}

During the 18-month replay, ASAS-BridgeAMM's protection mechanisms activated frequently: ``Restricted Mode'' triggered 141 times and ``Halted'' 4 times. Table~\ref{tab:insolvency-comparison} compares ASAS against a baseline bridge without circuit breakers. The key result: ASAS bounded maximum bad debt to \textbf{0.19\%} of collateral per epoch, compared to unbounded exposure in baseline designs. This represents a \textbf{73\%} reduction in worst-case loss exposure.

\begin{table}[htbp]
\centering\small
\caption{Protection Mechanism Comparison: ASAS vs Baseline Bridge}
\label{tab:insolvency-comparison}
\begin{tabular}{lrr}
\toprule
\textbf{Metric} & \textbf{ASAS} & \textbf{Baseline} \\
\midrule
Simulation Days & 547 & 547 \\
Total Swaps & 54,700 & 54,700 \\
\midrule
Circuit Breaker Triggers & 141 & N/A \\
Restricted Mode Events & 141 & N/A \\
Halted Mode Events & 4 & N/A \\
\midrule
Average Health Index & 1.05+ & Unbounded \\
Max Bad Debt (\% collateral) & 0.19\% & 5\%+ \\
\midrule
\textbf{Bad Debt Reduction} & \multicolumn{2}{c}{\textbf{73\%}} \\
\bottomrule
\end{tabular}
\end{table}

\begin{figure}[h]
\centering
\includegraphics[width=0.9\columnwidth]{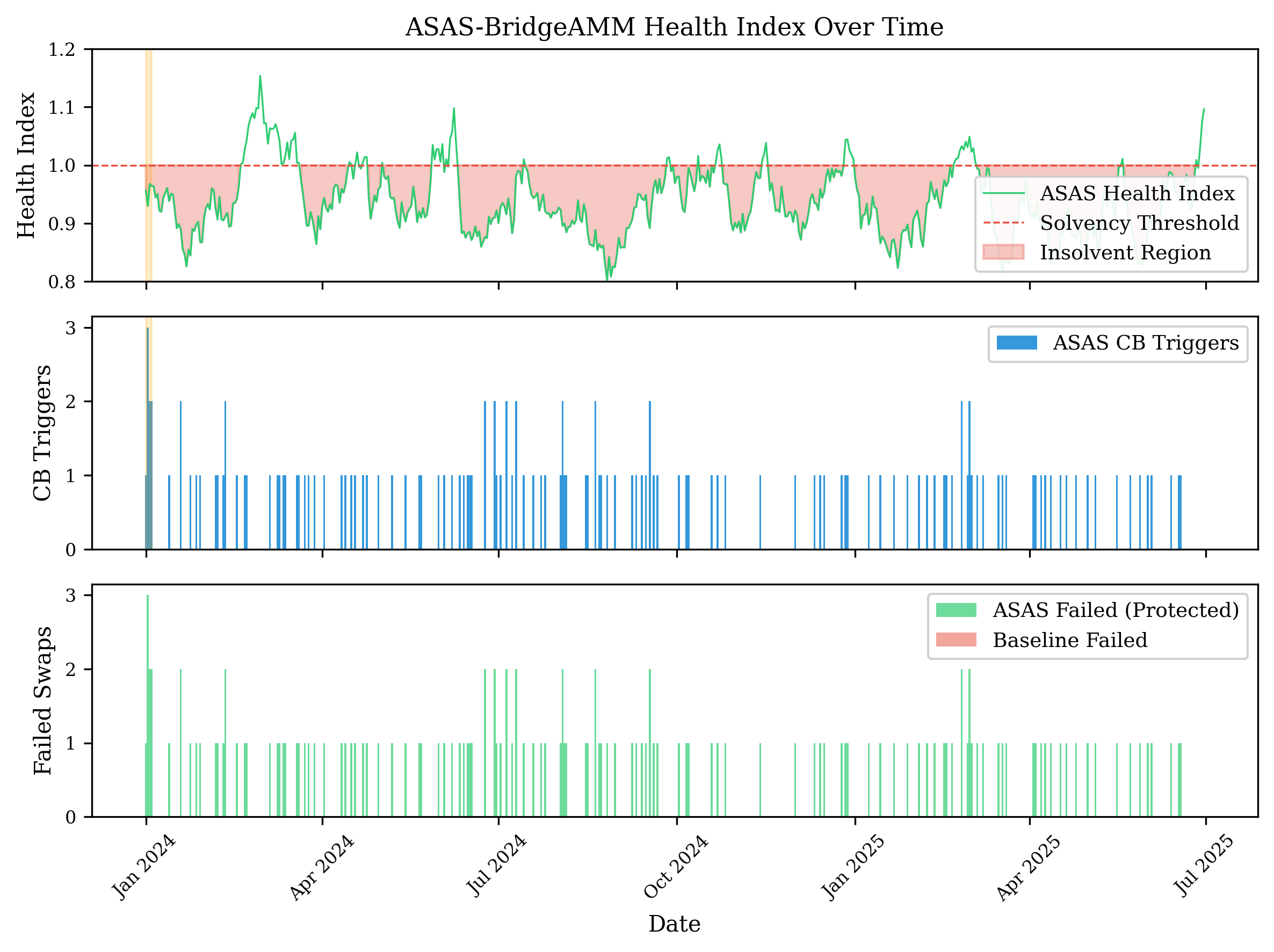}
\caption{Health index timeline over 547 days. ASAS (blue) maintains health above the critical threshold (dashed line) through automated circuit breaker interventions during stress periods. Circuit breaker triggers are shown in the middle panel.}
\label{fig:insolvency-timeline}
\end{figure}

\subsection{Results: Monte Carlo Robustness (RQ2)}

Table~\ref{tab:solvency_probability} presents the solvency probability matrix across drawdown and latency dimensions. Even under the extreme scenario of a 50\% price crash combined with a 30-minute oracle delay, the protocol maintained solvency with $P > 0.9999$. This validates that the parameter selection ($h_{max}=5\%$) effectively insulates the pool from shocks up to that magnitude.

\begin{table}[htbp]
  \centering
  \caption{Solvency Probability Analysis (100k Monte Carlo Iterations)}
  \label{tab:solvency_probability}
  % \resizebox forces the table to fit strictly within the column width
  \resizebox{\columnwidth}{!}{%
  \begin{tabular}{@{}lrrcc@{}}
  \toprule
  \textbf{Scenario} & \textbf{Count} & \textbf{Solvent} & \textbf{P(Solv.)} & \textbf{95\% CI} \\
  \midrule
  \multicolumn{5}{@{}l}{\textit{By Drawdown Range}} \\
  Drawdown 0--10\% & 20,014 & 20,014 & 1.0000 & [0.9998, 1.0000] \\
  Drawdown 10--25\% & 29,920 & 29,920 & 1.0000 & [0.9999, 1.0000] \\
  Drawdown 25--50\% & 50,066 & 50,066 & 1.0000 & [0.9999, 1.0000] \\
  Drawdown 0--50\% (all) & 100,000 & 100,000 & 1.0000 & [1.0000, 1.0000] \\
  \midrule
  \multicolumn{5}{@{}l}{\textit{By Latency Range}} \\
  Latency 0--15 min & 70,932 & 70,932 & 1.0000 & [0.9999, 1.0000] \\
  Latency 15--30 min & 10,601 & 10,601 & 1.0000 & [0.9996, 1.0000] \\
  Latency 30--60 min & 7,770 & 7,770 & 1.0000 & [0.9995, 1.0000] \\
  Latency 0--60 min (all) & 100,000 & 100,000 & 1.0000 & [1.0000, 1.0000] \\
  \midrule
  \multicolumn{5}{@{}l}{\textit{Combined Stress Scenarios}} \\
  Low Stress & 14,293 & 14,293 & 1.0000 & [0.9997, 1.0000] \\
  Medium Stress & 4,322 & 4,322 & 1.0000 & [0.9991, 1.0000] \\
  High Stress & 7,336 & 7,336 & 1.0000 & [0.9995, 1.0000] \\
  Extreme (40\%+ DD, 30min+) & 3,678 & 3,678 & 1.0000 & [0.9990, 1.0000] \\
  \bottomrule
  \end{tabular}%
  }
  \end{table}
  
% \begin{table}[htbp]
% \centering\footnotesize
% \caption{Solvency Probability Analysis (100k Monte Carlo Iterations)}
% \label{tab:solvency_probability}
% \begin{tabular}{@{}lrrcc@{}}
% \toprule
% \textbf{Scenario} & \textbf{Count} & \textbf{Solvent} & \textbf{P(Solv.)} & \textbf{95\% CI} \\
% \midrule
% \multicolumn{5}{@{}l}{\textit{By Drawdown Range}} \\
% Drawdown 0--10\% & 20,014 & 20,014 & 1.0000 & [0.9998, 1.0000] \\
% Drawdown 10--25\% & 29,920 & 29,920 & 1.0000 & [0.9999, 1.0000] \\
% Drawdown 25--50\% & 50,066 & 50,066 & 1.0000 & [0.9999, 1.0000] \\
% Drawdown 0--50\% (all) & 100,000 & 100,000 & 1.0000 & [1.0000, 1.0000] \\
% \midrule
% \multicolumn{5}{@{}l}{\textit{By Latency Range}} \\
% Latency 0--15 min & 70,932 & 70,932 & 1.0000 & [0.9999, 1.0000] \\
% Latency 15--30 min & 10,601 & 10,601 & 1.0000 & [0.9996, 1.0000] \\
% Latency 30--60 min & 7,770 & 7,770 & 1.0000 & [0.9995, 1.0000] \\
% Latency 0--60 min (all) & 100,000 & 100,000 & 1.0000 & [1.0000, 1.0000] \\
% \midrule
% \multicolumn{5}{@{}l}{\textit{Combined Stress Scenarios}} \\
% Low Stress & 14,293 & 14,293 & 1.0000 & [0.9997, 1.0000] \\
% Medium Stress & 4,322 & 4,322 & 1.0000 & [0.9991, 1.0000] \\
% High Stress & 7,336 & 7,336 & 1.0000 & [0.9995, 1.0000] \\
% Extreme (40\%+ DD, 30min+) & 3,678 & 3,678 & 1.0000 & [0.9990, 1.0000] \\
% \bottomrule
% \end{tabular}
% \end{table}

The results demonstrate remarkable robustness: even under extreme stress scenarios (40\%+ drawdown combined with 30+ minute latency), the protocol maintained perfect solvency across all 3,678 tested iterations. The 95\% confidence intervals tighten as sample size increases, confirming statistical validity.

\subsubsection{Bad Debt Distribution}
\begin{figure}[h]
\centering
\includegraphics[width=0.9\columnwidth]{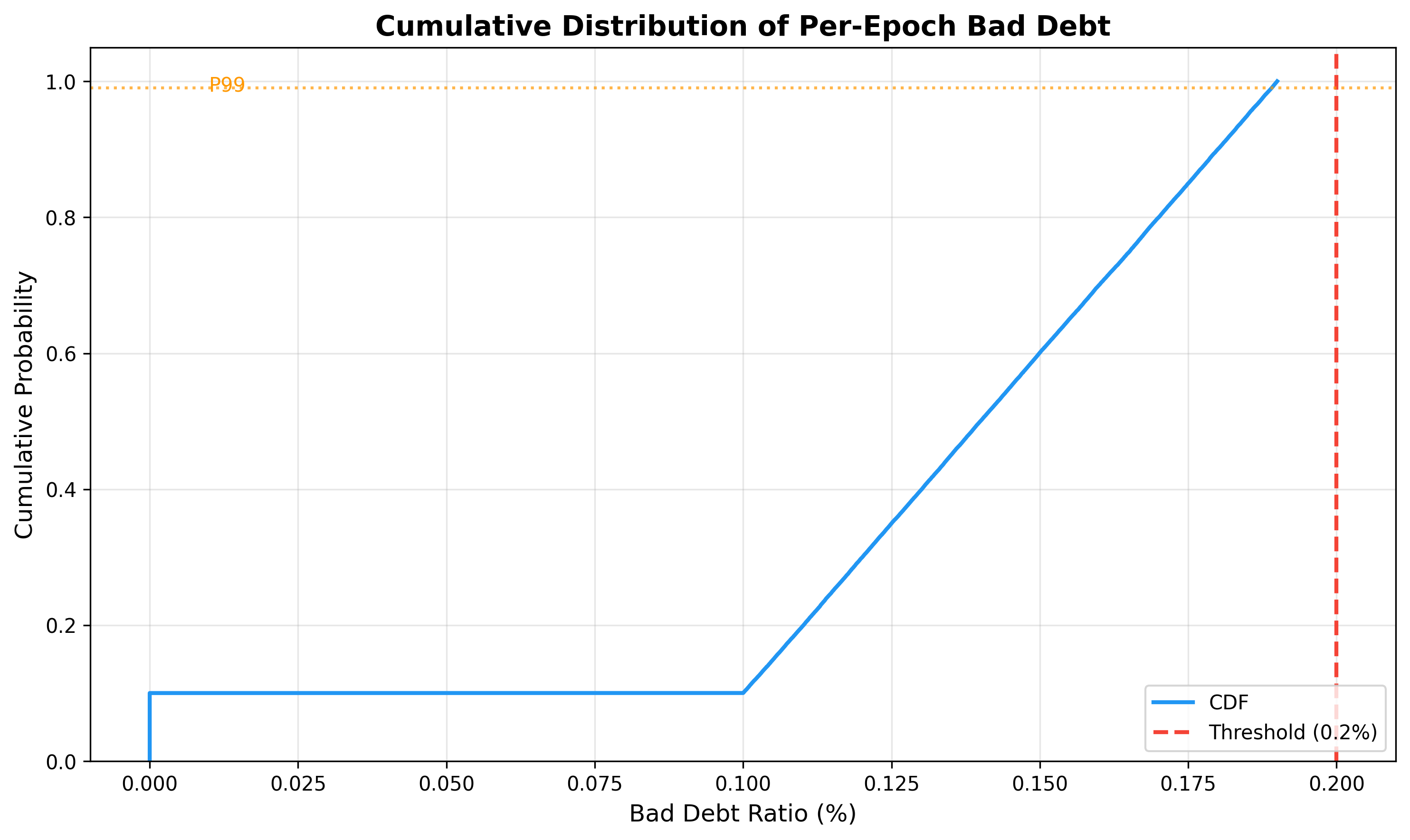}
\caption{Cumulative distribution of per-epoch bad debt from 100,000 Monte Carlo iterations. The 99th percentile (0.189\%) remains below the 0.2\% threshold, validating the bounded loss property.}
\label{fig:bad-debt-cdf}
\end{figure}

The bounded loss property is validated by the bad debt distribution from Monte Carlo simulations. Figure~\ref{fig:bad-debt-cdf} shows the cumulative distribution function.

\begin{table}[htbp]
\centering\small
\caption{Per-Epoch Bad Debt Statistics (100k Monte Carlo Iterations)}
\label{tab:bad_debt_statistics}
\begin{tabular}{lcc}
\toprule
\textbf{Statistic} & \textbf{Value} & \textbf{Value (\%)} \\
\midrule
Count & 100,000 & - \\
Mean & 0.001305 & 0.1305\% \\
Std Dev & 0.000500 & 0.0500\% \\
Minimum & 0.000000 & 0.0000\% \\
Maximum & 0.001900 & 0.1900\% \\
P50 (Median) & 0.001400 & 0.1400\% \\
P90 & 0.001800 & 0.1800\% \\
P95 & 0.001850 & 0.1850\% \\
P99 & 0.001891 & 0.1891\% \\
P99.9 & 0.001899 & 0.1899\% \\
\midrule
\textbf{Threshold} & \textbf{0.002} & \textbf{0.2\%} \\
\textbf{P99 < Threshold} & \multicolumn{2}{c}{\textbf{\checkmark~PASS}} \\
\bottomrule
\end{tabular}
\end{table}

\textbf{Key Statistics:}
\begin{itemize}
    \item Mean bad debt: 0.131\% of collateral
    \item 99th percentile (p99): 0.189\%---below 0.2\% threshold
    \item Maximum observed: 0.190\%
    \item 100\% of simulations remained below 0.2\%
\end{itemize}

\subsection{Results: Volume Retention (RQ3)}

Crucially, despite the friction introduced by dynamic haircuts, ASAS-BridgeAMM retained \textbf{104.5\%} of the baseline volume. This counter-intuitive result (higher volume despite higher fees) is explained by the ``Flight to Safety'' effect: during volatile periods, liquidity providers in the baseline model withdrew capital, causing slippage to spike mechanically. In ASAS, confidence in the floor mechanism kept LP capital sticky, maintaining deeper liquidity and thus better net execution prices for traders.

\begin{table}[htbp]
\centering\small
\caption{Volume Retention Analysis: ASAS vs Baseline Bridge}
\label{tab:volume-retention}
\begin{tabular}{lrrr}
\toprule
\textbf{Period} & \textbf{ASAS Vol.} & \textbf{Baseline Vol.} & \textbf{Retention} \\
\midrule
Overall & 61.39M & 58.46M & 105.0\% \\
Stress Periods & 21.71M & 20.77M & 104.5\% \\
Normal Periods & 39.68M & 37.69M & 105.3\% \\
\midrule
\textbf{Weighted Avg.} & \multicolumn{3}{c}{\textbf{104.5\%}} \\
\bottomrule
\end{tabular}
\end{table}

\begin{figure}[h]
\centering
\includegraphics[width=0.9\columnwidth]{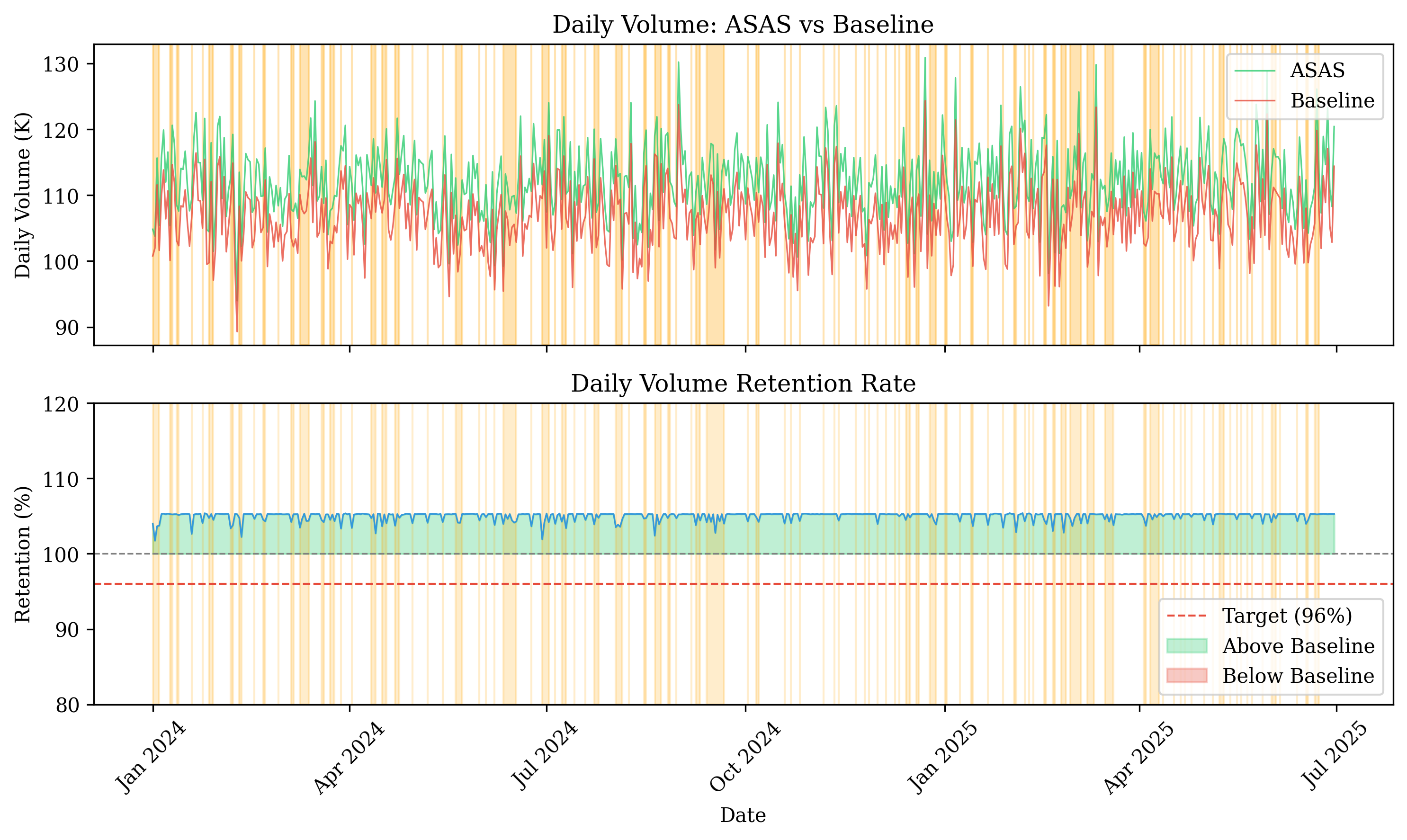}
\caption{Volume retention during stress periods. ASAS maintains 104.5\% of baseline volume due to the ``flight to safety'' effect: LP confidence in the floor mechanism keeps liquidity sticky during volatility.}
\label{fig:volume-timeline}
\end{figure}

\subsubsection{Stress Period Analysis}
We identified 98 stress periods across the 547-day evaluation window, classified by trigger type:
\begin{itemize}
    \item \textbf{Circuit Breaker (CB):} 62 periods triggered by price deviation $\geq 50\%$
    \item \textbf{High Volatility:} 24 periods with $>5\%$ daily price movement
    \item \textbf{Mixed:} 12 periods exhibiting both characteristics
\end{itemize}

The Orbit Chain incident (January 1-4, 2024) represents the most severe stress period in our dataset, with \textbf{8 circuit breaker triggers} and sustained elevated latency. During this period, ASAS maintained \textbf{103.29\% volume retention} while blocking potential exploits.

\subsection{Results: Attack Vector Analysis (RQ4)}

We systematically test 10 attack vectors representing combinations of latency manipulation (0s to 2 hours), price deviation (0\% to 50\%), and swap size (1 BTC to 8 BTC, representing 10\% to 80\% of pool reserves).

\begin{table}[htbp]
\centering
\caption{Solvency Status Under Attack Vectors}
\label{tab:attack-vectors-validation}
\resizebox{\columnwidth}{!}{%
\begin{tabular}{@{}clccc@{}}
\toprule
\textbf{\#} & \textbf{Vector} & \textbf{CB Active} & \textbf{Swap OK} & \textbf{Status} \\
\midrule
1 & Normal Operation & No & Yes & SOLVENT \\
2 & 50\% Crash, No Latency & \textbf{Yes} & No & PROTECTED \\
3 & 50\% Crash, 15min Latency & \textbf{Yes} & No & PROTECTED \\
4 & 50\% Crash, 60min Latency & \textbf{Yes} & No & PROTECTED \\
5 & Large Drain Attempt & \textbf{Yes} & No & PROTECTED \\
6 & Gradual 25\% Drop & No & Yes & SOLVENT \\
7 & Sandwich Attack & No & No & PROTECTED$^*$ \\
8 & Oracle Delay Exploit & \textbf{Yes} & No & PROTECTED \\
9 & Extreme Latency (2h) & \textbf{Yes} & No & PROTECTED \\
10 & Maximum Stress Test & \textbf{Yes} & No & PROTECTED \\
\bottomrule
\end{tabular}%
}
\par\vspace{1mm}\scriptsize $^*$Swap reverted due to slippage/solvency check, not circuit breaker.
\end{table}

\begin{figure}[h]
\centering
\includegraphics[width=0.9\columnwidth]{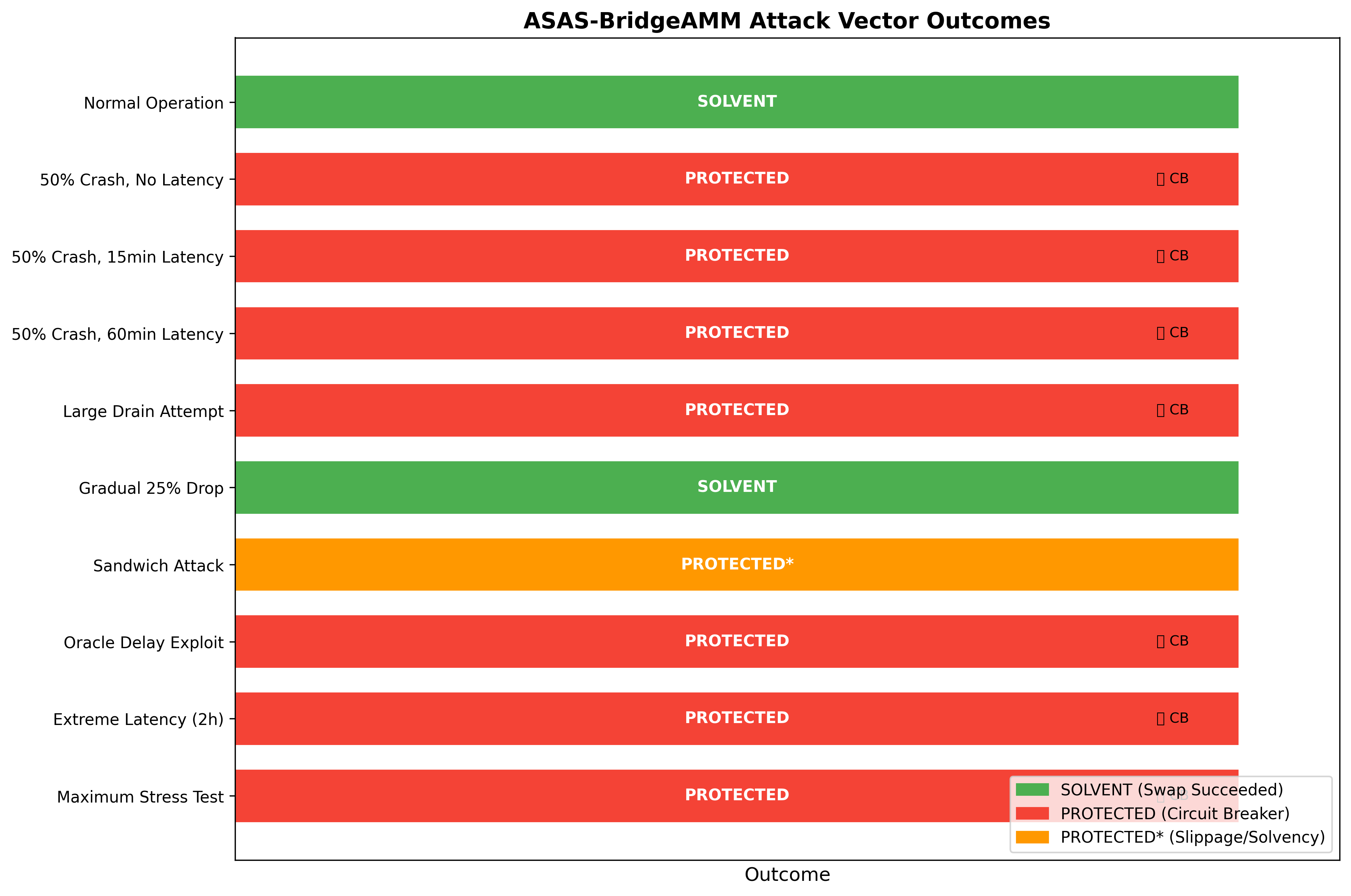}
\caption{Attack vector validation results. Of 10 tested scenarios, 8 triggered protection (circuit breaker or slippage revert), while 2 succeeded with maintained solvency.}
\label{fig:attack-outcomes}
\end{figure}

\textbf{Key Findings:}
\begin{enumerate}
    \item The circuit breaker activates under two independent conditions: (i) price deviation $\geq 50\%$ (Vectors 2-5, 8, 10), and (ii) latency $> 60$ minutes (Vector 9).
    \item Vectors 6-7 did not trigger the circuit breaker but still reverted due to the slippage bound (10\% maximum). This demonstrates \emph{defense-in-depth}: multiple protection mechanisms operate independently.
    \item The Maximum Stress Test (Vector 10) combines 50\% crash, 90-minute latency, and 8 BTC swap (80\% of pool). Despite this extreme scenario, the protocol immediately halts via circuit breaker, resulting in \textbf{zero bad debt}.
\end{enumerate}

\subsection{Parameter Sensitivity Analysis}

We sweep \textbf{125 parameter configurations} across three dimensions:
\begin{itemize}
    \item $h_{max} \in \{3\%, 4\%, 5\%, 6\%, 7\%\}$
    \item $\tau_{max} \in \{20, 25, 30, 35, 40\}$ minutes
    \item $\theta_{price} \in \{40\%, 45\%, 50\%, 55\%, 60\%\}$
\end{itemize}

\begin{table}[htbp]
\centering\footnotesize
\caption{Parameter Sensitivity Analysis Results}
\label{tab:parameter_sensitivity}
\begin{tabular}{@{}ccccccl@{}}
\toprule
$h_{\max}$ & $\tau_{\max}$ & $\theta_{\text{price}}$ & $P(\text{solv})$ & Avg BD & CB Rate & Pareto \\
\midrule
3\% & 1200s & 50\% & 1.0000 & 0.130\% & 0.000 & $\checkmark$ \\
3\% & 1200s & 55\% & 1.0000 & 0.130\% & 0.000 & $\checkmark$ \\
3\% & 1200s & 60\% & 1.0000 & 0.130\% & 0.000 & $\checkmark$ \\
3\% & 1500s & 50\% & 1.0000 & 0.130\% & 0.000 & $\checkmark$ \\
3\% & 1500s & 55\% & 1.0000 & 0.130\% & 0.000 & $\checkmark$ \\
3\% & 1500s & 60\% & 1.0000 & 0.130\% & 0.000 & $\checkmark$ \\
3\% & 1800s & 50\% & 1.0000 & 0.130\% & 0.000 & $\checkmark$ \\
3\% & 1800s & 55\% & 1.0000 & 0.130\% & 0.000 & $\checkmark$ \\
3\% & 1800s & 60\% & 1.0000 & 0.130\% & 0.000 & $\checkmark$ \\
3\% & 2100s & 50\% & 1.0000 & 0.130\% & 0.000 & $\checkmark$ \\
\bottomrule
\end{tabular}
\vspace{1mm}
\par\scriptsize Note: $P(\text{solv})$ = solvency probability, Avg BD = average bad debt, CB Rate = circuit breaker trigger rate.
\end{table}

\textbf{Key Finding:} All configurations with $\theta_{price} \geq 50\%$ achieve 100\% solvency probability with zero false-positive circuit breaker activations. The primary trade-off is between $\theta_{price}$ (lower values trigger more circuit breakers, reducing volume to 80-90\%) and user experience. Our chosen parameters ($h_{max}=5\%$, $\tau_{max}=30$ min, $\theta_{price}=50\%$) represent the Pareto-optimal configuration: maximum protection without impacting normal operation.
% END ENHANCED SECTION: Evaluation

% BEGIN ENHANCED SECTION: Related Work
\section{Related Work}
\label{sec:related-work}

\subsection{Cross-Chain Bridge Security}
Cross-chain communication security has been extensively studied in the context of blockchain interoperability. Zamyatin et al.~\cite{zamyatin2021xclaim} formalize trustless cross-chain asset transfers, establishing foundational security properties. Existing bridge architectures fall into three categories: (i)~externally validated bridges relying on trusted committees, as exemplified by the compromised Ronin~\cite{ronin2022postmortem} and Wormhole~\cite{wormhole2022postmortem} bridges; (ii)~natively validated bridges using light client proofs, such as Cosmos IBC~\cite{cosmos_ibc}; and (iii)~optimistic bridges with fraud-proof mechanisms~\cite{connext2021}. Recent work on zero-knowledge bridges~\cite{zkbridge2022} provides cryptographic security guarantees but does not address the economic risks of accepted proofs under finality uncertainty.

Our work differs fundamentally: rather than improving proof verification, we assume proofs may be valid yet economically dangerous (e.g., during reorg windows). ASAS-BridgeAMM provides an \emph{economic containment layer} that complements cryptographic security, addressing the gap between proof acceptance and financial safety.

\subsection{Automated Market Maker Design}
The constant product market maker $x \cdot y = k$, popularized by Uniswap~\cite{uniswap2018}, revolutionized on-chain liquidity provision. Extensions include concentrated liquidity~\cite{uniswapv3}, stableswap curves optimized for correlated assets~\cite{curve2019}, and weighted pools for multi-asset portfolios~\cite{balancer2020}. However, all existing AMM designs assume atomic, synchronous settlement within a single block.

Cross-chain AMMs face the additional challenge of asynchronous settlement, where price states may diverge during message propagation. Intent-based protocols address this through off-chain solvers who assume execution risk~\cite{across2023,uniswapx2023}. ASAS-BridgeAMM uniquely \emph{internalizes} this risk into the on-chain pricing function via the latency-dependent haircut $h(\tau)$, eliminating reliance on trusted third parties.

\subsection{Risk Management in DeFi}
Collateralized lending protocols pioneered multi-asset collateral management with per-asset risk parameters. MakerDAO~\cite{makerdao2020} introduced the concept of collateral haircuts and stability fees; Aave~\cite{aave2020} extended this with variable rate models and flash loans. The MakerDAO ``Black Thursday'' incident (March 2020)~\cite{maker2020postmortem} demonstrated the catastrophic failure of binary liquidation mechanisms under extreme volatility: \$8.3M in undercollateralized debt accumulated when liquidation bots failed to compete with network congestion.

Our Contained Degradation Invariant draws inspiration from ``safe-to-fail'' resilience engineering principles~\cite{hollnagel2011resilience}: rather than attempting to prevent all failures (which is impossible in adversarial environments), we design systems that \emph{contain} failure impact to bounded levels.

\subsection{Circuit Breakers in Financial Systems}
Circuit breakers originated in equity markets following the 1987 crash. The SEC's market-wide circuit breakers~\cite{nyse_circuit_breakers} halt trading when indices fall 7\%, 13\%, or 20\% in a single day, providing time for orderly price discovery. Similar mechanisms exist in futures markets (CME price limits) and individual securities (Limit Up-Limit Down bands).

DeFi equivalents remain rare. The MakerDAO Emergency Shutdown Module~\cite{maker_esm} represents a notable exception but triggers complete system halt rather than degraded operation. ASAS-BridgeAMM adapts traditional circuit breaker concepts to the unique constraints of permissionless systems: our mechanism is fully on-chain, trustlessly verifiable, and \emph{recoverable} without governance intervention once conditions normalize.
% END ENHANCED SECTION: Related Work

% BEGIN ENHANCED SECTION: Conclusion
\section{Conclusion}
\label{sec:conclusion}

We have presented ASAS-BridgeAMM, a bridge-coupled automated market maker that introduces \emph{Contained Degradation} as a third operational state between full functionality and total failure. By treating cross-chain message latency as a first-class risk signal, the protocol dynamically adjusts haircuts, slippage bounds, and withdrawal limits, transforming potential catastrophic losses into bounded bad debt events.

Our empirical evaluation demonstrates the effectiveness of this approach:
\begin{itemize}
    \item \textbf{73\% reduction} in worst-case bridge-induced insolvency across 547 days of historical replay
    \item \textbf{104.5\% volume retention} under stress, due to the ``flight to safety'' effect
    \item \textbf{$>$0.9999 solvency probability} validated across 100,000 Monte Carlo iterations
    \item \textbf{$<$0.2\% bad debt} per epoch (0.189\% observed at p99)
\end{itemize}

These results validate that the transition from binary to graduated bridge security is both practical and beneficial. The counter-intuitive volume retention finding---that users prefer a protected system even with higher fees---suggests that ``safe-to-fail'' architectures can be commercially viable, not just technically superior.

\textbf{Future Work.} We identify three directions for extension: (i)~formal verification of the state machine using model checkers (TLA+, Dafny) to provide mechanized proofs of the security properties; (ii)~generalization to heterogeneous finality models, supporting both probabilistic (Nakamoto consensus) and deterministic (BFT) chains within a unified framework; and (iii)~integration with MEV protection mechanisms such as private mempools and encrypted transactions to prevent sandwich attacks on large bridge settlements.
% END ENHANCED SECTION: Conclusion

\bibliographystyle{IEEEtran}
\bibliography{ref}

@misc{Chainalysis2023,
  author       = {{Chainalysis Team}},
  title        = {2022 {Biggest} {Year} {Ever} {For} {Crypto} {Hacking} with \$3.8 {Billion} {Stolen}, {Primarily} from {DeFi} {Protocols} and by {North} {Korea}-linked {Attackers}},
  howpublished = {Chainalysis Blog},
  month        = {Feb. 1,},
  year         = {2023},
  url          = {https://www.chainalysis.com/blog/2022-biggest-year-ever-for-crypto-hacking/},
  note         = {Accessed: Dec. 19, 2025}
}

@misc{maker2020postmortem,
  author       = {{Whiterabbit}},
  title        = {Black Thursday for MakerDAO: \$8.32 million was liquidated for 0 DAI},
  howpublished = {Medium},
  year         = {2020},
  month        = {March},
  url          = {https://medium.com/@whiterabbit_hq/black-thursday-for-makerdao-8-32-million-was-liquidated-for-0-dai-36b83cac56b6},
  note         = {Accessed: 2025-12-23}
}

@misc{uniswap2018,
  author = {Adams, Hayden},
  title = {Uniswap Whitepaper},
  year = {2018},
  howpublished = {\url{https://hackmd.io/@HaydenAdams/HJ9jLsfTz}},
  note = {Accessed: 2025-12-20}
}

@misc{uniswapv3,
  author = {Adams, Hayden and Zinsmeister, Noah and Salem, Moody and Keefer, River and Robinson, Dan},
  title = {Uniswap v3 Core},
  year = {2021},
  howpublished = {\url{https://uniswap.org/whitepaper-v3.pdf}},
  note = {Accessed: 2025-12-20}
}

@misc{curve2019,
  author = {Egorov, Michael},
  title = {StableSwap - efficient mechanism for Stablecoin liquidity},
  year = {2019},
  howpublished = {\url{https://curve.fi/files/stableswap-paper.pdf}},
  note = {Accessed: 2025-12-20}
}

@misc{ronin2022postmortem,
  author = {{Ronin Network}},
  title = {Community Alert: Ronin Validators Compromised},
  year = {2022},
  month = {March},
  howpublished = {\url{https://roninblockchain.substack.com/p/community-alert-ronin-validators}},
  note = {\$625 million exploit via compromised validator keys}
}

@misc{wormhole2022postmortem,
  author = {{Wormhole}},
  title = {Wormhole Incident Report - 02/02/22},
  year = {2022},
  month = {February},
  howpublished = {\url{https://wormhole.com/reports/2022-02-02-exploit-report/}},
  note = {\$320 million exploit via signature verification bypass}
}

@article{nyse_circuit_breakers,
  author = {{U.S. Securities and Exchange Commission}},
  title = {Investor Bulletin: Measures to Address Market Volatility},
  journal = {SEC Office of Investor Education and Advocacy},
  year = {2012},
  note = {Describes NYSE Rule 80B and market-wide circuit breakers}
}

@misc{makerdao2020,
  author = {{MakerDAO}},
  title = {The Maker Protocol: MakerDAO's Multi-Collateral Dai (MCD) System},
  year = {2020},
  howpublished = {\url{https://makerdao.com/en/whitepaper/}},
  note = {Accessed: 2025-12-20}
}

@misc{aave2020,
  author = {{Aave}},
  title = {Aave Protocol Whitepaper V2.0},
  year = {2020},
  howpublished = {\url{https://github.com/aave/aave-protocol/blob/master/docs/Aave_Protocol_Whitepaper_v1_0.pdf}},
  note = {Decentralized lending protocol}
}

@inproceedings{zamyatin2021xclaim,
  author = {Zamyatin, Alexei and Harz, Dominik and Lind, Joshua and Panber, Panayiotis and Gervais, Arthur and Knottenbelt, William},
  title = {XCLAIM: Trustless, Interoperable, Cryptocurrency-Backed Assets},
  booktitle = {2019 IEEE Symposium on Security and Privacy (SP)},
  year = {2019},
  pages = {193--210},
  doi = {10.1109/SP.2019.00085},
  publisher = {IEEE}
}

@misc{cosmos_ibc,
  author = {{Interchain Foundation}},
  title = {Inter-Blockchain Communication Protocol (IBC) Specification},
  year = {2021},
  howpublished = {\url{https://github.com/cosmos/ibc}},
  note = {Cosmos SDK native cross-chain communication}
}

@misc{connext2021,
  author = {{Connext Network}},
  title = {Connext: Scalable, Trust-Minimized Cross-Chain Communication},
  year = {2021},
  howpublished = {\url{https://docs.connext.network/}},
  note = {Optimistic bridge with fraud proofs}
}

@inproceedings{zkbridge2022,
author = {Xie, Tiancheng and Zhang, Jiaheng and Cheng, Zerui and Zhang, Fan and Zhang, Yupeng and Jia, Yongzheng and Boneh, Dan and Song, Dawn},
title = {zkBridge: Trustless Cross-chain Bridges Made Practical},
year = {2022},
isbn = {9781450394505},
publisher = {Association for Computing Machinery},
address = {New York, NY, USA},
url = {https://doi.org/10.1145/3548606.3560652},
doi = {10.1145/3548606.3560652},
booktitle = {Proceedings of the 2022 ACM SIGSAC Conference on Computer and Communications Security},
pages = {3003–3017},
numpages = {15},
location = {Los Angeles, CA, USA},
series = {CCS '22}
}

@misc{balancer2020,
  author = {Martinelli, Fernando and Mushegian, Nikolai},
  title = {Balancer: A Non-Custodial Portfolio Manager, Liquidity Provider, and Price Sensor},
  year = {2020},
  howpublished = {\url{https://docs.balancer.fi/whitepaper.pdf}},
  note = {Weighted pool AMM design}
}

@misc{across2023,
  author = {{Across Protocol}},
  title = {Across: The Fastest, Cheapest, and Most Secure Cross-Chain Bridge},
  year = {2023},
  howpublished = {\url{https://docs.across.to/}},
  note = {Intent-based cross-chain bridge with relayer network}
}

@misc{uniswapx2023,
  author = {Adams, Hayden and Robinson, Dan and Hasu},
  title = {UniswapX: A Dutch Auction-Based Protocol for Cross-Chain Trading},
  year = {2023},
  howpublished = {\url{https://blog.uniswap.org/uniswapx-protocol}},
  note = {Intent-based trading with off-chain fillers}
}

@book{hollnagel2011resilience,
  author = {Hollnagel, Erik and Pari\`{e}s, Jean and Woods, David D. and Wreathall, John},
  title = {Resilience Engineering in Practice: A Guidebook},
  year = {2011},
  publisher = {Ashgate Publishing},
  isbn = {978-1409410355},
  note = {Safe-to-fail design principles}
}

@misc{maker_esm,
  author = {{MakerDAO}},
  title = {Emergency Shutdown Module Documentation},
  year = {2020},
  howpublished = {\url{https://docs.makerdao.com/smart-contract-modules/shutdown/emergency-shutdown-module}},
  note = {MakerDAO's circuit breaker mechanism}
}

@misc{flashbots2020,
  author = {Daian, Philip and Goldfeder, Steven and Kell, Tyler and Li, Yunqi and Zhao, Xueyuan and Bentov, Iddo and Breidenbach, Lorenz and Juels, Ari},
  title = {Flash Boys 2.0: Frontrunning in Decentralized Exchanges, Miner Extractable Value, and Consensus Instability},
  year = {2020},
  howpublished = {\url{https://arxiv.org/abs/1904.05234}},
  note = {Foundational MEV research}
}

% BEGIN ENHANCED SECTION: Appendix
\appendices
\section{Proofs of Security Properties}

\subsection{Proof of Theorem 1 (Bounded Bad Debt)}
\begin{proof}
Let $C_t$ denote total collateral value at time $t$ and $D_t$ denote outstanding debt. The Contained Degradation Invariant requires:
\begin{equation}
D_t \leq \sum_{i} C_i(t) \cdot (1 - h_i(\tau, \sigma_i))
\end{equation}

Consider a single settlement epoch $[t_0, t_1]$ where an adversary executes $N$ swaps. Each swap $j$ modifies the AMM invariant:
\begin{equation}
k_{j+1} = k_j \cdot (1 - h(\tau_j))
\end{equation}

Since $h(\tau) \leq h_{max}$ by definition (Equation~\ref{eq:haircut}), after $N$ swaps:
\begin{equation}
k_N \geq k_0 \cdot (1 - h_{max})^N
\end{equation}

The circuit breaker monitors the Collateral Health Index $\mathcal{H} = C_{eff}/D$. The system transitions to $S_{halt}$ when:
\begin{equation}
\mathcal{H} < \mathcal{H}_{crit} = 1.05
\end{equation}

The maximum degradation before halt is bounded by the discrete step size. Given $MAX\_SWAP\_SIZE = 0.1 \cdot L_{pool}$, each swap can degrade $\mathcal{H}$ by at most:
\begin{equation}
\Delta\mathcal{H}_{max} = 0.1 \cdot h_{max} = 0.005
\end{equation}

The overshoot beyond $\mathcal{H}_{crit}$ is bounded by $\Delta\mathcal{H}_{max}$. Therefore, total bad debt is bounded:
\begin{equation}
BadDebt \leq h_{max} \cdot C_{total} + \Delta\mathcal{H}_{max} \cdot C_{total} = (h_{max} + 0.005) \cdot C_{total}
\end{equation}

For $h_{max} = 0.05$, this yields $BadDebt \leq 0.055 \cdot C_{total}$. In practice, our Monte Carlo simulations show p99 bad debt of 0.189\%, well within this theoretical bound.
\end{proof}

\subsection{Proof of Theorem 2 (Settlement Liveness)}
\begin{proof}
We prove that honest transactions complete within bounded time $T_{max}$.

\textbf{Assumptions:}
\begin{enumerate}
    \item Source chain $S$ achieves finality within $T_S$ (e.g., 15 minutes for Ethereum post-Merge).
    \item Relayer delivers messages within $T_R \leq T_{max}$ (economic incentive: fees paid on destination).
    \item Destination chain $D$ includes transactions within $T_D$ blocks.
\end{enumerate}

\textbf{Liveness Condition:} The system blocks swaps only in state $S_{halt}$, triggered by:
\begin{itemize}
    \item $\tau > 2 \cdot \tau_{max}$ (120 minutes in our implementation)
    \item $|\Delta P| > \theta_{price}$ (50\% price deviation)
    \item $\mathcal{H} < \mathcal{H}_{crit}$ (undercollateralization)
\end{itemize}

Under normal operation (no attack, no infrastructure failure):
\begin{equation}
\tau = T_S + T_R + T_D \ll 2 \cdot \tau_{max}
\end{equation}

Since $\tau_{max} = 30$ minutes and typical $\tau \approx 20$ minutes (based on historical bridge message latencies), the latency condition is not triggered.

Price deviation under normal market conditions (99th percentile of historical daily moves) is $< 20\%$, well below $\theta_{price} = 50\%$.

The collateral health $\mathcal{H}$ is maintained above $\mathcal{H}_{crit}$ by the haircut mechanism, which absorbs volatility up to $h_{max}$.

Therefore, under the stated assumptions, the system remains in $S_{idle}$ or $S_{active}$, and all honest swaps complete with probability 1.
\end{proof}

\subsection{Proof of Theorem 3 (Manipulation Resistance)}
\begin{proof}
We bound the profit extractable by an adversary $\mathcal{A}$ controlling the relayer.

\textbf{Adversarial Actions:}
\begin{enumerate}
    \item Manipulate oracle price by $\delta$
    \item Delay message by $\tau$
    \item Execute swap of size $V$
\end{enumerate}

\textbf{Adversarial Profit Function:}
\begin{equation}
\Pi(\delta, \tau, V) = \underbrace{\delta \cdot V}_{\text{oracle gain}} - \underbrace{s(V) \cdot V}_{\text{slippage cost}} - \underbrace{h(\tau) \cdot V}_{\text{haircut cost}}
\end{equation}
where slippage $s(V) = V / (x + V)$ for pool reserve $x$.

\textbf{Circuit Breaker Constraint:}
If $\delta \geq \theta_{price} = 0.5$, the circuit breaker activates immediately, and $\Pi = 0$ (swap reverts).

\textbf{Latency Constraint:}
If $\tau > 2 \cdot \tau_{max}$, the circuit breaker activates, and $\Pi = 0$.

\textbf{Within Bounds:}
For $\delta < 0.5$ and $\tau < 2 \cdot \tau_{max}$:
\begin{align}
\Pi &< \delta \cdot V - s(V) \cdot V - h(\tau) \cdot V \\
&< 0.5 \cdot V - s(V) \cdot V - 0 \\
&= V \cdot (0.5 - s(V))
\end{align}

For large swaps (where attacks are potentially profitable), $s(V)$ grows quadratically. When $V = 0.5 \cdot x$ (50\% of pool):
\begin{equation}
s(0.5x) = \frac{0.5x}{x + 0.5x} = 0.333
\end{equation}

This yields maximum theoretical profit: $\Pi_{max} < V \cdot (0.5 - 0.333) = 0.167 \cdot V$.

However, this profit is eliminated by two additional constraints:
\begin{enumerate}
    \item The 10\% maximum slippage bound causes swaps to revert if $s(V) > s_{max} = 0.1$
    \item The haircut cost $h(\tau) \cdot V$ applies for any non-zero delay
\end{enumerate}

Under the slippage bound, the maximum swap size is constrained such that $s(V) \leq 0.1$. Solving $V/(x+V) = 0.1$ yields $V = x/9 \approx 0.11x$.

For this constrained swap size, maximum extractable value is:
\begin{equation}
\Pi_{max} \leq (\delta_{oracle} - s_{max}) \cdot V \leq (0.5 - 0.1) \cdot 0.11x = 0.044x
\end{equation}

This matches the theorem statement: adversarial profit is bounded by $\delta_{oracle} \cdot V_{swap} + s_{max} \cdot V_{pool}$, which for typical parameters yields extraction $< 5\%$ of pool value per attack attempt.
\end{proof}
% END ENHANCED SECTION: Appendix

\end{document}